\documentclass[11pt,a4paper]{article}
\usepackage{epsf,feynmf,amsmath}
\usepackage{graphicx,color,mathrsfs,amsmath,amssymb}
\newcommand{\cd}{\makebox[0.08cm]{$\cdot$}}
\newcommand{\reels}{\mathbb{R}}

\newcommand{\sla}{\not\!}
\title{\bf The fate of infrared divergences in a finite formulation of field theory: QED revisited}
\author{Jean-Fran\c cois Mathiot \thanks{e-mail: jean-francois.mathiot@clermont.in2p3.fr} \vspace{0.3cm}\\
{\small \em  Universit\'e Clermont Auvergne, Laboratoire de Physique de Clermont,
} \\ {\small \em CNRS/IN2P3, BP10448, F-63000 Clermont-Ferrand, France}}
\begin{document}
\maketitle
\bibliographystyle{unsrt}
\abstract{
Within the framework of the recently proposed Taylor-Lagrange regularization procedure, we reanalyze the calculation of radiative corrections in $QED$ at next to leading order. Starting from a well defined local bare Lagrangian, the use of this regularization procedure enables us to manipulate fully finite elementary amplitudes in the ultra-violet as well as infra-red regimes, in physical $D=4$ space-time dimensions and for physical massless photons, as required by gauge invariance. We can thus separately calculate the electromagnetic form factors of the electron and the cross-section for real photon emission, each quantity being finite in these physical conditions.  We then discuss the renormalization group equations within this regularization procedure. Thanks to the taming of infra-red divergencies,  the renormalization group equation associated to the (physical) effective charge exhibits an ultra-violet stable fixed point at $\alpha^*=0$, showing an asymptotic freedom type behavior. We finally consider the case of two mass scales, one low and one heavy, paying particular attention to the natural decoupling properties between heavy and light degrees-of-freedom. As a direct consequence, the fine structure constant should be zero in the limit of massless electrons.
}
%
%%%%%%%%%%%%%%%%%%%%%%%%%%%%%%%%%%%%%%%%%%%%%%%%
\section{Introduction} \label{introduction}
Following the recent development of a regularization procedure based on the nature of quantum fields as operator valued distributions ($OPVD$) - the so-called Taylor-Lagrange regularization scheme ($TLRS$) \cite{GW_mass,GW} - we shall consider in this study the case of quantum electrodynamics ($QED$) in the one-loop, next to leading order ($NLO$), approximation.   
This regularization procedure originates from the observation that the divergences of bare amplitudes can be traced back to the violation of causality due to the ill-defined product of distributions at the same point \cite{bogol,bosh,stuec, wigh,haag,aste} (see  also Refs.~\cite{tkach,schw}). Since the Lagrangian we start from is constructed from the product of fields or derivative of fields at the same point - it is thus called {\it local} - the calculation of any elementary amplitude must  be done with great care.
The correct mathematical treatment for such a case is  known since a long time \cite{EG, GB,schar}. More recently, these considerations led to the construction of $TLRS$. According to this procedure, physical fields are constructed as $OPVD$, these distributions being applied on test functions with well defined mathematical properties. Since this scheme is completely finite - in a sense that will be defined below -  it is not plagued by any arbitrariness due to the way divergences in the ultra-violet ($UV$) as well as infra-red ($IR$) regimes are cancelled. We can therefore concentrate on the most important, physical, consequences of the finite renormalization of the bare amplitudes, as in any interacting many-body system. \\

The main properties of $TLRS$ can be characterized by the following  two essential features:
\begin{itemize}
\item $TLRS$ enables us to give a well defined meaning to the Lagrangian we start from. This is of course also the case using dimensional regularization ($DR$). Both regularization procedures are  thus called {\it a-priori} in contrast to {\it a-posteriori} regularization procedures, like for instance using a na\" ive cut-off in momentum space. In this latter case, the regularization is done a posteriori at the level of each elementary amplitude and not at the level of the Lagrangian itself.
\item The calculation of any elementary amplitude in $TLRS$ is done in physical conditions, {\it i.e.} in four space-time dimensions, with no additional non-physical degrees of freedom like for instance (infinitely massive) Pauli-Villars fields, and for massless photons. $TLRS$ is thus called an {\it intrinsic} regularization procedure, in contrast to $DR$ for which  elementary amplitudes still diverge in physical conditions. This last procedure is thus called {\it extrinsic}.
\end{itemize}

The construction of $TLRS$ enables us to treat at the same time $UV$ as well as $IR$ singular operators \cite{GW_mass,GW}. The  $IR$ singularities do occur in particular when massless degrees of freedom are involved. A textbook example of such a case is given by one-loop corrections in  $QED$. Using $DR$ for instance, such calculation requires to give, in intermediate steps, a small non-zero mass to the photon. The subsequent massless limit  is then taken at the very end of the calculation. These  $IR$ singularities  should be properly taken care of before any physical consequences can be drawn from the calculation of a given physical observable. We shall investigate in our study the physical consequences  of using $TLRS$ for the calculation of $NLO$ corrections in $QED$. Although these corrections are by now well known examples, the use of this regularization procedure enables a completely new analysis, free from any $UV$ and $IR$ divergences. This implies in particular that no intermediate renormalization is necessary. In this sense, $TLRS$ is at the same time a regularization procedure and a renormalization scheme, with the same acronym. The only renormalization we should worry about is the field strength renormalization for external, on-shell, particles. Thanks to the lack of any $IR$ divergences, this renormalization factor is well defined for massless photons. \\

The behavior of  any elementary amplitude and any physical observable is governed by two arbitrary scales, as already explained in Ref.~\cite{jfm}:
\begin{itemize}
\item  the {\it regularization scale}  denoted  by $\eta$.  It is inherent to the regularization procedure which is used to give a mathematical well defined meaning to the local bare Lagrangian we start from. {\it This scaling variable is dimensionless in $TLRS$}. 
\item The energy scale $M$ at which an experiment is performed in order to fix the value of the parameters of the Lagrangian. It is more precisely a set of scales, like for instance in $\phi^4$ theories. There is however only one scale in the case of $QED$. We call this scale the {\it renormalization point} since it fixes the kinematical condition where the finite (physical) renormalization of the bare parameters is performed. 
\end{itemize}
These two arbitrary scales are the ones relevant for the calculation of the running of the two universal coupling constants - the bare and the physical ones - using the renormalization group ($RG$) equations \cite{jfm}. The bare coupling constant depends on the regularization scale $\eta$ only and is denoted by $\alpha_0(\eta)$, while the physical one depends on the renormalization point $M$ only and is denoted by $\alpha_M(M)$\footnote{When $\alpha_M(M)$ is not directly accessible from an experiment at any value of $M$ - as it is the case for $QED$ - we can consider equivalently an effective charge directly related to a physical observable as we shall see in Sec.~\ref{ff}.}. These two coupling constants are universal in the sense that they can be identified independently of the choice of any regularization procedure or any renormalization scheme. Moreover, they  can  be defined both in the perturbative as well as non-perturbative regimes. Note that the calculation of the physical coupling constant $\alpha_M$, at $M\ne 0$,  is only made possible when $IR$ divergences are properly taken care of, as we shall see in Sec.~\ref{phys}.

We would like to emphasize  the very different nature of these two coupling constants. On the one hand, the bare one - $\alpha_0(\eta)$ - is defined at the level of the  bare Lagrangian, and knows nothing about the renormalization scheme which will be used, if any, nor about the physical state which is realized in Nature, like for instance  in the presence of spontaneous symmetry breaking. On the other hand, the physical coupling constant $\alpha_M(M)$ is a definite  property of this physical state, and is independent of the regularization procedure which has been used. 
The running of these two coupling constants is therefore governed by two separate $RG$ equations. The one associated to the $\eta$-dependence of $\alpha_0$, called $RGE(\eta)$, is mass-independent since it is associated to the local character of the Lagrangian we start from, {\it i.e.} to the $UV$ limit of elementary amplitudes in momentum space. The $RG$ equation associated to the $M$-dependence of $\alpha_M$, called $RGE(M)$, is mass-dependent since the kinematical condition $M$ is finite and not  arbitrarily large. 

Once elementary amplitudes have been calculated, like for instance the self-energy of the electron, the polarization operator of the photon and the electromagnetic vertex correction, one should consider physical observables. Apart of course from the physical  mass of the electron which is used to fix its bare mass, or the fine structure constant which is used to fix the bare coupling constant and its $\eta$-dependence, the first simple non-trivial observable is the elastic $e^--e^\pm$  scattering. As a direct consequence of the unique properties of $TLRS$ recalled above, we shall see in Sec.~\ref{phys} - in the one-photon-exchange approximation - that this scattering amplitude is finite in physical conditions {\it i.e.} with a massless photon.  We shall also check  that the cross-section for soft-photon bremstrahlung is  finite in these conditions. The calculation of the electromagnetic form factors of the electron - for an arbitrary precision of the experimental apparatus in order to separate real photon emission from virtual vertex corrections - is thus made possible for the first time. As we shall see in Sec.~\ref{RGE}, this has a non trivial consequence in the high energy limit. In this limit, $QED$  exhibits an $UV$ stable fixed point for the effective charge with $\alpha^*=0$, showing an asymptotic freedom type behavior.\\

 The plan of our article is the following. We calculate in Sec.~\ref{elm} the elementary amplitudes in $QED$ at $NLO$, and check the Ward identities. The electromagnetic form factors of the electron together with soft-photon bremstrahlung are calculated in Sec.~\ref{phys}. We discuss in Sec.~\ref{RGE} the use of  the $RG$ equations as well as  the case of two mass scales and the limit of massless electrons. Our conclusions are drawn in Sec.~\ref{conc}. We recall  in  \ref{TLRS} the main physical properties of $TLRS$, while the  calculation of all relevant integrals is detailed in \ref{integral}.

 %%%%%%%%%%%%%%%%%%%%%%%%%%%%%%%%%%%%%%%%%%%%%%%%
\section{Elementary amplitudes}\label{elm}
For illustration purposes on how to use $TLRS$ in practice, we recall in this section the calculation of the elementary amplitudes in $QED$ at $NLO$. For simplicity, we restrict ourself to the Feynman gauge. The use of different gauges is discussed in  Ref.~\cite{GW_gauge}. All the necessary integrals are detailed in \ref{integral}. 
\subsection{Self-energy of the electron} \label{selfsec}
The electron self-energy is written, with the appropriate test functions $f_\sigma$ (see \ref{TLRS}),
\begin{equation} \label{self}
\Sigma(p)=-ie^2\lim_{\sigma \to1^-} \int \frac{d^4k}{(2\pi)^4} \frac{\gamma^\mu (\sla p \ - \sla k+m) \gamma_\mu }{k^2[(p-k)^2-m^2]} 
f_\sigma\left[\frac{k^2}{m^2}\right] f_\sigma\left[\frac{(p-k)^2}{m^2}\right],
\end{equation}
where $e$ is the physical  charge and $m$ the physical mass of the electron.
The $IR$ singularity at $k=0$ is taken  care of by the first  test function. We thus get 
\begin{equation}
\Sigma(p)=2ie^2 \left[(\sla p -2m)\overline I_0-\gamma_\nu \overline I_1^\nu\right],
\end{equation}
where the integrals $\overline I_0$ and $\overline I_1^\mu$ are given by
\begin{equation}\label{J1m}
(\overline I_0,\overline I_1^\nu)(p)=\lim_{\sigma \to1^-}\int \frac{d^4 k}{(2\pi)^4} \frac{(1,k^\nu)}{ k^2[(p-k)^2+m^2]}
f_\sigma\left[\frac{k^2}{m^2}\right] f_\sigma\left[\frac{(k-p)^2}{m^2}\right].
\end{equation}
They are calculated in \ref{integral}. With the most general decomposition
\begin{equation} \label{Bdef}
\Sigma(p)=m A(p^2)+\sla p B(p^2),
\end{equation}
we have 
\begin{subequations} \label{AB}
\begin{eqnarray}
A(p^2)&=&\frac{\alpha}{\pi} \int_0^1dx\ \left[ \mbox{Log} \frac{\eta^2}{\Delta_p}+cte\right], \\
B(p^2)&=&-\frac{\alpha}{2\pi} \int_0^1dx (1-x)\ \left[ \mbox{Log} \frac{\eta^2}{\Delta_p}+cte\right],
\end{eqnarray}
\end{subequations}
with 
\begin{equation}
\Delta_p=1-\frac{p^2}{m^2}(1-x),
\end{equation}
and $\alpha=\frac{e^2}{4\pi}$.  In the above equations, and in all this study, we have indicated by $cte$ a constant term, independent of any kinematical variable, in order to remind us  that the regularization scale $\eta$ is defined up to a multiplicative constant (see \ref{TLRS}). Note that the integrals in Eq.~(\ref{AB}) do not involve any test function anymore since $A$ and $B$ are finite. We recover here the standard result \cite{peski}.

We shall also need  in Sec.~\ref{phys}  the electron field strength renormalization factor $Z$. This factor is written as \cite{peski}, at $NLO$,
\begin{equation} \label{Z2}
Z=1+\left.\frac{d\Sigma}{d\!\!\sla p} \right \vert_{\not p = m}=2m^2\left[A'(m^2)+B'(m^2)\right]+B(m^2).
\end{equation}
The calculation of $A'$ and $B'$ requires some care since both quantities involve $IR$ singular operators \cite{GW_gauge}.  They are calculated in \ref{integral}. We thus get
\begin{equation} \label{Z2b}
Z\equiv1+\delta=1-\frac{\alpha}{4\pi} \left[ \mbox{Log}\ \eta^2+cte \right].
\end{equation}
This factor is free from any $IR$ divergences although it is calculated with a massless photon.

\subsection{Vacuum polarization of the photon}
The calculation of the polarization operator of the photon proceeds similarly. We have
\begin{equation} \label{Pi}
\Pi^{\mu\nu}(q)=ie^2\lim_{\sigma \to1^-} \int \frac{d^4k}{(2\pi)^4} \frac{Tr[\gamma^\mu (\sla k+m) \ \gamma^\nu (\sla k - \sla q+m)]}{(k^2-m^2)[(k-q)^2-m^2]} 
 f_\sigma\left[\frac{k^2}{m^2}\right] f_\sigma\left[\frac{(k-q)^2}{m^2}\right].
\end{equation}
We thus get
\begin{equation} \label{pol}
\Pi^{\mu\nu}(q)=4ie^2\left[ 2\overline J_2^{\mu \nu} - g^{\mu \nu} \overline J_2 + g^{\mu \nu} \overline J_1^\rho q_\rho - q^\mu \overline J_1^\nu - q^\nu \overline J_1^\mu +m^2 g^{\mu \nu} \overline J_0\right].
\end{equation}
The various integrals entering in Eq.~(\ref{pol}) are given  by
\begin{multline}
(\overline J_0,\overline J_1^\mu,\overline J_2,\overline J_2^{\mu\nu})(q)=\lim_{\sigma \to1^-}\int \frac{d^4k}{(2\pi)^4} \frac{(1,k^\mu,k^2,k^\mu k^\nu)}{(k^2-m^2)[(k-q)^2-m^2]} \\
\times f_\sigma\left[\frac{k^2}{m^2}\right]f_\sigma\left[\frac{(k-q)^2}{m^2} \right].
\end{multline}
They are calculated in \ref{integral}. We  have finally 
\begin{equation} \label{pseudo}
\Pi^{\mu\nu}(q^2)= \Pi(q^2) \left[g^{\mu \nu} q^2-q^\mu q^\nu \right],
\end{equation}
where
\begin{equation} \label{labelpi}
\Pi(q^2)=-\frac{2\alpha}{\pi} \int_0^1 dx\ x(1-x)\left[\mbox{Log}\frac{\eta^2}{\Delta_q}+cte\right],
\end{equation}
with 
\begin{equation} \label{Delta}
\Delta_q=1+\frac{Q^2}{m^2}x(1-x),
\end{equation}
and $Q^2=-q^2$. We can check explicitly here that the photon propagator remains transverse as required by gauge invariance. It is instructive to calculate the limiting cases $Q^2\ll m^2$ and $Q^2\gg m^2$. We get
\begin{subequations}
\begin{eqnarray}
\Pi(Q^2\ll m^2)&=&-\frac{\alpha}{3\pi}\left[\mbox{Log}\ \eta^2-\frac{Q^2}{5 m^2}+cte\right],\\
\Pi(Q^2\gg m^2)&=&\frac{\alpha}{3\pi}\mbox{Log}\frac{Q^2}{m^2}.
\end{eqnarray}
\end{subequations}
These results will be used in Sec.~\ref{phys} for the calculation of the electromagnetic form factors of the electron.

\subsection{Electromagnetic vertex} \label{elmvertex}
For simplicity, we calculate here the electromagnetic vertex for external on-shell electrons only. It is given, at $NLO$, by 
\begin{multline}
\Lambda^{\mu}(p,q)=-ie^2 \bar u(p')\lim_{\sigma \to1^-} \int \frac{d^4k}{(2\pi)^4} \frac{\gamma^\rho (\sla{p'} - \sla k+m) \gamma^{\mu}(\sla{p} - \sla k+m) \gamma_{\rho}}{k^2[(p'-k)^2-m^2][(p-k)^2-m^2]}  \\
\times  f_\sigma\left[\frac{k^2}{m^2}\right] f_\sigma\left[\frac{(p-k)^2}{m^2}\right] f_\sigma\left[\frac{p'-k)^2}{m^2}\right]u(p),
\end{multline}
where $\bar u(p')$ and $u(p)$ are the  Dirac spinors, and $q=p'-p$.
As usual, we  decompose the electromagnetic vertex into two parts. The first one, denoted by $\Lambda_{UV}^\mu$, is a divergent contribution in the $UV$ domain in the absence of test functions, while the second one, denoted by $\Lambda_{IR}^\mu$, is convergent in this domain but has still $IR$ divergences which should be properly taken care of. The first one depends explicitly on the regularization scale $\eta$ while the second one does not.  We get 
\begin{multline}
\Lambda^{\mu}_{UV}(p,q)=-ie^2 \bar u(p')\lim_{\sigma \to1^-} \int \frac{d^4k}{(2\pi)^4} \frac{\gamma^\rho \sla{k} \gamma^{\mu}\sla k \gamma_{\rho}}{k^2[(p'-k)^2-m^2][(p-k)^2-m^2]} \\
\times f_\sigma\left[\frac{k^2}{m^2}\right] f_\sigma\left[\frac{(p-k)^2}{m^2}\right] f_\sigma\left[\frac{(p'-k)^2}{m^2}\right] u(p).
\end{multline}
We thus can write
\begin{equation}
\Lambda^{\mu}_{UV}(p,q)=2ie^2\bar u(p')\left[2 \overline K_2^{\mu \nu}\gamma_\nu  - \overline K_2 \gamma^\mu\right]u(p),
\end{equation}
where the integrals $\overline K_2$ and $\overline K_2^{\mu\nu}$ are given by
\begin{multline}
(\overline K_2,\overline K_2^{\mu\nu})(p,q)= 
\lim_{\sigma \to1^-}\int \frac{d^4k}{(2\pi)^4} \frac{(k^2,k^{\mu}k^{\nu})}{k^2[(p'-k)^2-m^2][(p-k)^2-m^2]}\\
\times f_\sigma\left[\frac{k^2}{m^2}\right] f_\sigma\left[\frac{(p-k)^2}{m^2}\right] f_\sigma\left[\frac{(p'-k)^2}{m^2}\right].
\end{multline}
They are calculated in  \ref{integral}. We  finally have
\begin{equation}
\Lambda^{\mu}_{UV}(p,q)= \Phi_1^{UV}(Q^2)\ \bar u(p')\gamma^\mu u(p) , \label{gamma}
\end{equation}
with
\begin{equation}
\Phi_1^{UV}(Q^2)=\frac{\alpha}{2\pi} \int_0^1dx\int_0^{1-x} dy\left[\mbox{Log}\frac{\eta^2}{\Delta}+cte\right],
\end{equation}
and $\Delta=(x+y)^2+\frac{Q^2}{m^2}xy$. By a standard change of variable \cite{peski} with $w=x+y$ and $y=w \xi$, we get, after integration over $w$ and with the change of notation $\xi \to x$,
\begin{equation} \label{wxi}
\Phi_1^{UV}(Q^2)=\frac{\alpha}{4\pi} \int_0^1dx \left[\mbox{Log}\frac{\eta^2}{\Delta_q}+cte\right],
\end{equation}
with $\Delta_q$ given in Eq.~(\ref{Delta}).
In the limits $Q^2\ll m^2$ and $Q^2\gg m^2$ we have
\begin{subequations}
\begin{eqnarray}
\Phi_1^{UV}(Q^2\ll m^2)&=&\frac{\alpha}{4\pi}\left[\mbox{Log}\ \eta^2-\frac{Q^2}{6 m^2}+cte\right],\\
\Phi_1^{UV}(Q^2\gg m^2)&=&-\frac{\alpha}{4\pi}\mbox{Log}\frac{Q^2}{m^2}.
\end{eqnarray}
\end{subequations}
These results will be used in Sec.~\ref{phys} for the calculation of the electromagnetic form factors of the electron.

The contribution $\Lambda^{\mu}_{IR}$ is finite in the $UV$ domain but has still singularities in the $IR$ domain in the absence of test functions, as well known. We can write, using the on-shell conditions for the external legs,
\begin{multline}
\Lambda^{\mu}_{IR}(p,q)=-4ie^2\bar u(p')\left[\left[(p+p')^\mu\gamma_\nu-(p+p')_\nu\gamma^\mu\right]\overline K_1^\nu-m \overline K_1^\mu\right.\\
+\left.\gamma^\mu\left(m^2+\frac{Q^2}{2}\right)\overline K_0\right]u(p),
\end{multline}
where the integrals $\overline K_0$ and $\overline K_1^\lambda$ are  given by
\begin{multline}
(\overline K_0,\overline K_1^\lambda)(p,q)= 
\lim_{\sigma \to1^-}\int \frac{d^4k}{(2\pi)^4} \frac{(1,k^\lambda)}{k^2[(p'-k)^2-m^2][(p-k)^2-m^2]}\\
\times f_\sigma\left[\frac{k^2}{m^2}\right] f_\sigma\left[\frac{(p-k)^2}{m^2}\right] f_\sigma\left[\frac{(p'-k)^2}{m^2}\right].
\end{multline}
They are calculated in \ref{integral}. This gives
\begin{equation}
\Lambda^{\mu}_{IR}(p,q)=  \bar u(p')\left[ \gamma^\mu \Phi_1^{IR} + \frac{i}{2m} \sigma^{\mu\nu} q_\nu\ \Phi_2\right] u(p),
\end{equation}
with
\begin{subequations}
\begin{eqnarray}
\Phi_1^{IR}(Q^2)&=&\frac{\alpha}{2\pi}\lim_{\sigma \to1^-}\int_0^1dx \int_0^{1-x} dy\frac{1}{\Delta} \left[\frac{}{}\left[(x+y)^2+2(x+y)-2\right]\right.\nonumber\\
&&+\left.\frac{Q^2}{m^2} \left[(x+y-xy-1)\right]\right]F_\sigma,\\
\Phi_2(Q^2)&=&-\frac{\alpha}{\pi}\int_0^1dx \int_0^{1-x} dy \frac{1}{\Delta} \left[(x+y)(x+y-1)\right].
\end{eqnarray}
\end{subequations}
We have kept in $\Phi_1^{IR}$  the relevant test functions, summarized by $F_\sigma$, in order to take care of the $IR$ singularities. We recover here of course the well known result for $\Phi_2(Q^2)$ since it has no infrared divergences. The continuum limit $\sigma \to1^-$ can then be taken immediately in this case. We thus have, using the results of \ref{IRdiv} and with the same change of variables as above,
\begin{subequations}
\begin{eqnarray}
\Phi_1^{IR}(Q^2)&=&\frac{\alpha}{4\pi}\int_0^1\frac{dx}{\Delta_q}\left[5-2\mbox{Log}\Delta_q
+\frac{Q^2}{m^2} \left[2-x(1-x)-\mbox{Log}\Delta_q\right] \right]\label{phiIR},\\
\Phi_2(Q^2)&=&\frac{\alpha}{2\pi}\int_0^1  \frac{dx}{\Delta_q} .
\end{eqnarray}
\end{subequations}
In the particular limits of very small or very large momentum transfer, we have
\begin{subequations}
\begin{eqnarray}
\Phi_1^{IR}(Q^2\ll m^2)&=&\frac{\alpha}{4\pi}\left(5 + \frac{2Q^2}{3m^2}\right) \label{IRinf},\\
\Phi_2(Q^2\ll m^2)&=&\frac{\alpha}{2\pi}\left(1-\frac{Q^2}{6m^2}\right),
\end{eqnarray}
\end{subequations}
and
\begin{subequations}
\begin{eqnarray}
\Phi_1^{IR}(Q^2\gg m^2)&=&-\frac{\alpha}{4\pi}\left[\mbox{Log} \frac{Q^2}{m^2}\right]^2,\\
\Phi_2(Q^2\gg m^2)&=&\frac{\alpha}{\pi} \frac{m^2}{Q^2}\mbox {Log} \frac{Q^2}{m^2}.
\end{eqnarray}
\end{subequations}
This completes the calculation of the electromagnetic vertex in $QED$, using $TLRS$. As expected, all expressions are finite in physical conditions, {\it i.e.} in four space-time dimensions and with a massless photon. Note the $\left[\mbox{Log} \frac{Q^2}{m^2}\right]^2$ behavior of $\Phi_1^{IR}$ in the large $Q^2$ limit. We shall come back to this point in the next Sections.

\subsection{Ward-Takahashi identity} \label{ward}
With our notations, the Ward-Takahashi identity is written as
\begin{equation} \label{WT}
\Lambda^\mu(p,0)=-\bar u(p) \left[ \frac{\partial}{\partial p_\mu}\Sigma(p)\right] u(p).
\end{equation}
From the expression (\ref{self}) for $\Sigma(p)$ we have
\begin{eqnarray}\label{WT2}
\bar u(p) \left[ \frac{\partial}{\partial p_\mu}\Sigma(p)\right] u(p)&=&- \Lambda^\mu (p,0) \\
&&-4ie^2 \bar u(p) (\gamma^\mu p_\nu -p^\mu \gamma_\nu)\overline u(p) K_1^\nu (p,p) \nonumber \\
&&-ie^2\bar u(p) \left[\lim_{\sigma \to1^-} \int \frac{d^4k}{(2\pi)^4} \frac{\gamma^\mu (\sla p \ - \sla k+m) \gamma_\mu }{k^2[(p-k)^2-m^2]} \frac{\partial}{\partial p_\mu} F_\sigma\right] u(p),\nonumber
\end{eqnarray}
with the integral $\overline K_1^\nu$ calculated in \ref{integral}. With the on-shell condition, the second term in the r.h.s. of Eq.~(\ref{WT2}) is identically zero. Moreover, the third term in the r.h.s. of this equation is also zero in the continuum limit since the derivative of the test functions is zero everywhere except in the asymptotic region where it goes to zero more rapidly than any power of the momentum as a rapidly decreasing function (see \ref{TLRS}).
This insures the conservation of the Ward identities at that order. Note that this is only true in the continuum limit.

%%%%%%%%%%%%%%%%%%%%%%%%%%%%%%%%%%%%%%%%%%%%%%%%
\section{Physical observables} \label{phys}
From the above results, it is easy to anticipate that the use of $TLRS$ enables us to calculate, for the first time, physical observables free from any $IR$ divergences. We shall concentrate in this first study on the electromagnetic form factors of the electron. Within $TLRS$, these form factors are unambiguous since they are $IR$ finite and do not depend on any regularization scale. They can be extracted from a combination of $e^--e^-$ and $e^--e^+$ elastic scattering cross-sections.
It is commonly said that  $IR$ divergences in the calculation of these cross-sections are cancelled once soft-photon bremstrahlung - which is not separated out experimentally below a given energy  threshold $\Delta E$ of the photon - is considered (the well known Bloch-Nordsieck mechanism \cite{block}). Note that, strictly speaking, these $IR$ divergences  in first order perturbation theory have not disappeared anyhow in this case but  they have just been reinterpreted in terms of  $IR$ divergences when  $\Delta E$ tends to $0$ in any {\it Gedanken} experiment. This is indeed the price to pay, when using $DR$ for instance, for not having treated properly these $IR$ divergences from the start, in physical conditions. 

We shall show in Sec.~\ref{real}  how the use of $TLRS$ enables us to calculate the cross-section for the emission of real massless photons. We first concentrate on the calculation of the electromagnetic form factors.

\subsection{The electromagnetic form factors of the electron} \label{ff}
\begin{figure}
\centerline{\includegraphics[width=22pc]{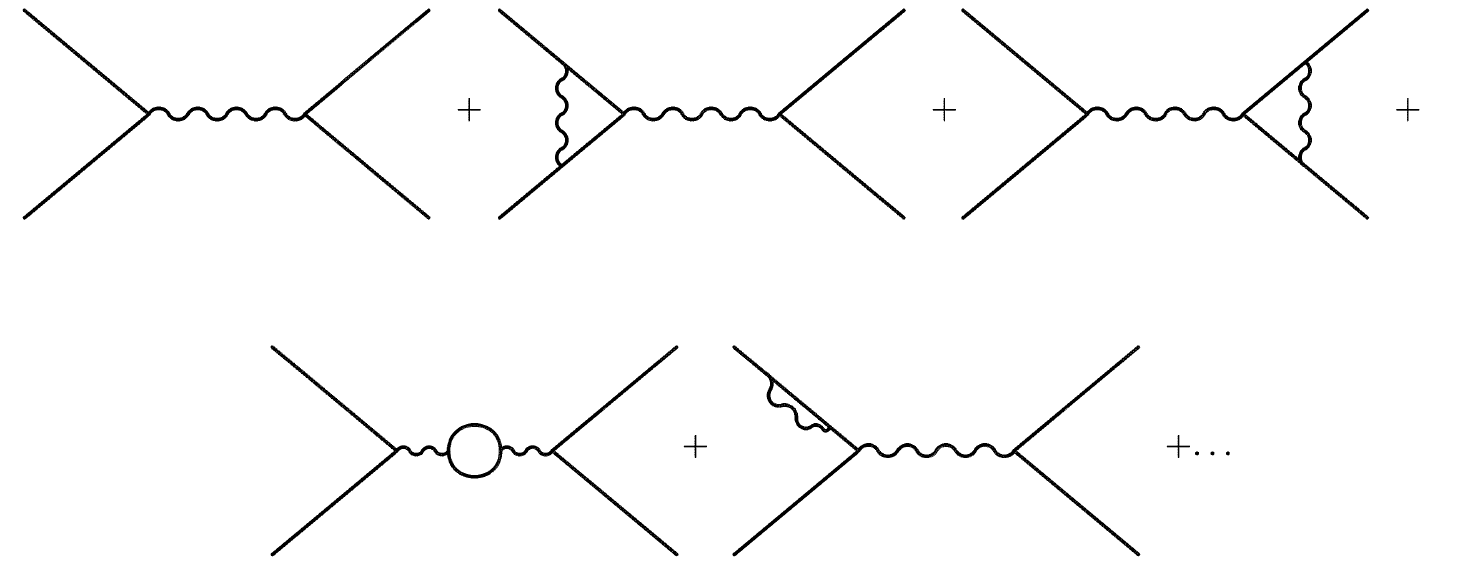}}
\begin{center}
\caption{Elementary $e-e$ amplitude at $NLO$ in the one-photon exchange approximation. The dots indicate similar contributions with self-energy corrections on any of the external legs.  \label{fig}}
\end{center}
\end{figure}

The physical amplitude for elastic $e^--e^-$ scattering is written as, in the Feynman gauge and in the one-photon exchange approximation,
\begin{equation}
{\cal M}_{ee}=i \frac{e^2}{q^2}\left[ \bar u(p'_1)\Gamma_\mu u(p_1)\right] \times \left[\bar u(p'_2)\Gamma^\mu u(p_2)\right],
\end{equation}
with
\begin{equation}
\Gamma_\mu=\gamma_\mu F_1(Q^2)+\frac{i}{2m}\sigma_{\mu \nu}q^\nu F_2(Q^2).
\end{equation}
The  form factor $F_1$ is normalized to $F_1(Q^2=0)=1$ by definition of the physical electric charge $e$ of the electron. 
The various contributions to ${\cal M}_{ee}$ at $NLO$ are indicated on Fig.~\ref{fig} in the one-photon-exchange approximation.
All these contributions have been calculated in Sec.~\ref{elm}. We thus get, in terms of the bare coupling constant $\alpha_0$,
\begin{equation} \label{form}
\alpha F_1^2(Q^2)=\alpha_0 Z^2+\alpha\Pi(Q^2)+2\alpha\Phi_1(Q^2),
\end{equation}
with $\Phi_1=\Phi_1^{UV}+\Phi_1^{IR}$, while $F_2$ is simply identical to $\Phi_2$ at that order.
The value of $\alpha_0$, and its $\eta$-dependence, is then fixed from the calculation of the fine structure constant at $Q^2=0$, with
\begin{equation}
\alpha_0=\alpha\left[1-\Pi(0)-2\delta-2\Phi_1(0)\right],
\end{equation}
where $\delta$ is defined in Eq.~(\ref{Z2b}). We can then calculate the  form factor $F_1$ of the electron, with
\begin{equation} \label{F1Q}
F_1(Q^2)=1+\frac{1}{2}\left[\Pi(Q^2)-\Pi(0)\right]+\left[\Phi_1(Q^2)-\Phi_1(0)\right].
\end{equation}
While this expression is of course not new and refers to the early days of $QED$ \cite{Schwi}, it is  calculated here with a massless photon, as demanded by gauge invariance, and is free from any $IR$ divergences. As expected $F_1(Q^2)$ is  $\eta$-independent, as it should. 

It is instructive to calculate $F_1(Q^2)$ in the two limiting kinematical conditions $Q^2 \ll m^2$ and $Q^2 \gg m^2$. We get immediately, from the results of Sec.~\ref{elm},
\begin{subequations} \label{ffas}
\begin{eqnarray}
F_1(Q^2\ll m^2)&=&1+\frac{\alpha}{4\pi}\frac{Q^2}{m^2}\frac{19}{30}+ {\cal{O}}(\alpha^2),\\
F_1(Q^2\gg m^2)&=&1-\frac{\alpha}{4\pi} \left[\mbox{Log} \frac{Q^2}{m^2}\right]^2+ {\cal{O}}(\alpha^2).
\end{eqnarray}
\end{subequations}
The value of $F_1$ in the large momentum region is of particular interest. While it shows the usual $\left[ \mbox{Log}\ Q^2\right]^2$ behavior, this contribution is  finite although the calculation is done, from the start, with a massless photon. This is a direct consequence of using $TLRS$ which enables us to tame both $UV$ and $IR$ divergences in physical conditions. 

\subsection{Soft photon emission} \label{real}
The elementary cross-section for the emission of a single soft photon  is well known \cite{ster}. It is given, in first order perturbation theory,  by 
\begin{equation}
d\sigma(p\to p'+\gamma)=d\sigma(p\to p') I,
\end{equation}
with
\begin{equation} \label{soft}
I=e^2\lim_{\sigma \to 1^-}\int\frac{d^3{\bf k}}{(2\pi)^3 2\omega}\left[\frac{2p\cd p'}{p\cd k\ p'\cd k} -\frac{m^2}{(p\cd k)^2}-\frac{m^2}{(p'\cd k)^2}\right] f_\sigma \left[\frac{\omega^2}{m^2}\right],
\end{equation}
where $\omega=\vert {\bf k}\vert$. We have kept explicitly in Eq.~(\ref{soft}) the appropriate test function for the outgoing photon \cite{GW_gauge}. For large momentum transfer, and with an upper limit $\Delta E$ for the energy of the outgoing real photon, we have
\begin{equation}
I=\frac{2\alpha}{\pi} \mbox{Log}\frac{Q^2}{m^2}\lim_{\sigma \to 1^-}\int_0^{\Delta E} \frac{d\omega}{\omega}f_\sigma\left[\frac{\omega^2}{m^2}\right]\equiv \frac{2\alpha}{\pi} \mbox{Log}\frac{Q^2}{m^2} J.
\end{equation}
With $X=\omega^2/m^2$, we can write
\begin{equation}
J=\frac{1}{2}\lim_{\sigma \to 1^-}\int_0^{\frac{(\Delta E)^2}{m^2}} \frac{dX}{X}f_\sigma(X).
\end{equation}
From the properties of $TLRS$, and by definition of the Pseudo-function  (see \ref{TLRS}), we get
\begin{equation}
J=\frac{1}{2}\int_0^{\frac{(\Delta E)^2}{m^2}} dX Pf\left[\frac{1}{X}\right]=\frac{1}{2}\mbox{Log}\left[\frac{(\Delta E)^2}{m^2}\right].
\end{equation}
The elementary cross-section for the emission of a soft-photon is thus 
\begin{equation} \label{softIR}
d\sigma(p\to p'+\gamma)=d\sigma(p\to p')\ \frac{\alpha}{\pi}\ \mbox{Log}\frac{Q^2}{m^2}\ \mbox{Log}\left[\frac{(\Delta E)^2}{m^2}\right].
\end{equation}
This cross-section does not show  any $IR$ divergences associated to the zero mass of the photon anymore. One can thus treat separately virtual corrections to the electromagnetic form factors of the electron from the emission of soft real photons. This contribution should further be summed up to all orders in order to account for the emission of an arbitrary number of real photons \cite{ster}. This gives the usual Sudakov-type correction, for large $Q^2$, with
\begin{equation}\label{suda}
I\to \mbox{exp}[I]=\mbox{exp}\left[-\frac{\alpha}{\pi}\ \mbox{Log}\frac{Q^2}{m^2}\ \mbox{Log}\left(\frac{m^2}{(\Delta E)^2}\right)\right].
\end{equation}
This correction tends to $0$ when $\Delta E$ gets very small, leaving only the virtual  photon contribution embedded in the electromagnetic form factors of the electron, independently of the ability of the experimental apparatus in disantangling the emission of soft real photons from elastic $e-e$ scattering.

\subsection{ Comparison with dimensional regularization}
It is particularly interesting to compare our results with those using $DR$ for instance, as far as $IR$ divergences are concerned. In this latter approach, the only meaningful contribution to consider in order to get an $IR$ finite physical observable is the sum of the $IR$ divergent contributions for both the virtual vertex correction  and the contribution from (non-detected) soft-photon emission below a given photon energy $\Delta E$. This sum is simply {given \cite{lifs,bere}},  for the differential cross-section at large $Q^2$, by:
\begin{equation} \label{softDR}
d\sigma(p\to p')\ \frac{2\alpha}{\pi}\ \mbox{Log}\frac{Q^2}{m^2}\left[ -\mbox{Log}\frac{m}{\lambda}+\mbox{Log}\frac{2\Delta E}{\lambda}\right],
\end{equation}
where $\lambda$ is a small finite mass of the photon. In $TLRS$, the only ($IR$-finite) contribution to compare with comes from soft-photon emission, as calculated in Eq.~(\ref{softIR}). It is rewarding to check that both contributions (\ref{softIR}) and (\ref{softDR}) are identical for large values of $\Delta E/m$. This insures that the use of $TLRS$ should be compatible, at $NLO$ at least, with all $QED$-related experimental results already analyzed in the framework of $DR$. In all these calculations, $\Delta E$ should be fixed from the exact threshold for (non-detected) soft-photon emission according to the characteristics of each experimental apparatus. Contrarily to $DR$, the use of $TLRS$ enables us to unambiguously define the electromagnetic form factors of the electron independently of any experimental considerations, as expected from general arguments for a well-defined theoretical framework. As a direct consequence, we can calculate the effective charge of the electron for an arbitrary value of the energy scale, as we shall see below.

%%%%%%%%%%%%%%%%%%%%%%%%%%%%%%%%%%%%%%%%%%%%%%%%
\section{The renormalization group equations} \label{RGE}
\subsection{Decoupling equation}
As already mentioned in the Introduction, we must consider separately two $RG$ equations. The first one, $RGE(\eta)$, is associated to the independence of physical observables on the dimensionless regularization scale $\eta$. It concerns the bare parameters and is mass-independent. The second one, $RGE(M)$, is associated to the independence of physical observables on the dimensionful renormalization point $M$. It concerns the effective charge $\alpha_M$ and is mass-dependent. 
These two $RG$ equations can be obtained simultaneously from the calculation of the physical coupling constant, or effective charge, in terms of the bare one, and similarly for the bare mass. By definition of $\alpha_M$, we can write
\begin{equation} \label{alphaM}
\alpha_M(M)\equiv \alpha F_1^2(Q^2=M^2)=\alpha_0(\eta) Z(\eta)^2+\alpha\Pi(\eta,M^2)+2\alpha \Phi_1(\eta,M^2),
\end{equation}
with $Z$ given by Eq.~(\ref{Z2}). For completeness, we have  indicated in the above equation the various $\eta$- and $M$-dependences. From the results of Sec.~\ref{elm}, we can see that the $\eta$-dependence of $\Pi(\eta,M^2)$ and $\Phi_1(\eta,M^2)$ can be easily separated out, with
\begin{subequations} \label{piphi}
\begin{eqnarray}
\Pi(\eta,M^2)&=&-\frac{\alpha}{3\pi}\left( \mbox{Log}\ \eta^2+ cte\right)+\overline \Pi(M^2),\\
\Phi_1(\eta,M^2)&=&\frac{\alpha}{4\pi}\left( \mbox{Log}\ \eta^2+cte\right)+\overline \Phi_1^{UV}(M^2)+\overline \Phi_1^{IR}(M^2), \label{Phi1}
\end{eqnarray}
\end{subequations}
and
\begin{subequations}
\begin{eqnarray} 
\overline \Pi(M^2)&=&\frac{2\alpha}{\pi} \int_0^1 dx\ x(1-x)\mbox{Log}\left[1+\frac{M^2}{m^2}x(1-x)\right],\\
\overline \Phi_1^{UV}(M^2)&=&-\frac{\alpha}{4\pi} \int_0^1dx\ \mbox{Log}\left[1+\frac{M^2}{m^2}x(1-x)\right] \label{overphi1}.
\end{eqnarray}
\end{subequations}
We have defined in these equations  $\overline \Phi_1^{IR}=\Phi_1^{IR}(M^2)-\Phi_1^{IR}(M^2=0)$ with $\Phi_1^{IR}$ given in Eq.~(\ref{phiIR}). For convenience, we have normalized $\overline \Phi_1^{IR}$ to $\overline \Phi_1^{IR}(M^2=0)=0$ by including $\Phi_1^{IR}(M^2=0)$ into the constant $cte$ in Eq.~(\ref{Phi1}). This simply corresponds to a finite on-mass-shell renormalization condition.
We can thus write Eq.~(\ref{alphaM}) as
\begin{equation}\label{decou}
\alpha_M(M)-\alpha\overline \Pi(M^2)-2\alpha\overline \Phi_1^{UV}(M^2)-2\alpha \overline  \Phi_1^{IR}(M^2)= 
\alpha_0(\eta)-\frac{\alpha^2}{3\pi}\left(\mbox{Log}\ \eta^2+cte\right)\equiv \alpha.
\end{equation}
Thanks to the Ward identity, the $\eta$-dependence of $\alpha_0(\eta)$ is given only by the vacuum polarization of the photon, as well known, while the energy scale dependence of the physical coupling constant includes in addition the contribution from the electromagnetic vertex. As shown in Ref.\cite{jfm}, this decoupling property persists when these radiative corrections are summed up to all orders.

Note that the common value of Eq.~(\ref{decou}) - which should be independent of both $\eta$ and $M$ -  is just the fine structure constant $\alpha$ since it is by construction the value of $\alpha_M(M=0)$. With the results of Sec.~\ref{elm}, we thus have in both limits $M\ll m$ and $M\gg m$,
\begin{subequations}
\begin{eqnarray}
\alpha_M(M\ll m)&=&\alpha+\frac{\alpha^2}{4\pi} \frac{19}{15}\frac{M^2}{m^2}+ {\cal{O}} (\alpha^3),\\
\alpha_M(M\gg m)&=&-\frac{\alpha^2}{2\pi} \left(\mbox{Log}\frac{M^2}{m^2}\right)^2+ {\cal{O}} (\alpha^3).
\end{eqnarray}
\end{subequations}

\subsection{$\beta$ functions} \label{betaf}
The decoupling equation (\ref{decou}) is instructive from many points of view.

{\it i)} The behavior of $\alpha_0$ as a function of $\eta$ should be compared with the behavior of $\alpha_R(\mu)$ in $DR$ in the minimal subtraction scheme ($MS$) as a function of the unit of mass \cite{thoo} $\mu$ of $DR$. They both give the same mass-independent $\beta$ function with, in $TLRS$,
\begin{equation} \label{betaeta}
\beta_\eta \equiv \eta \frac{\partial \alpha_0(\eta)}{\partial \eta} = \frac{2\alpha_0^2}{3\pi}+{\cal O}(\alpha_0^3).
\end{equation}
This behavior should not be identified with any physical pattern. It is just the remnant of the scaling properties associated to the local character of the Lagrangian we start from, independently of the relevance of this Lagrangian to describe  the physical reality in a given energy domain.

{\it ii)} The behavior of the physical coupling constant $\alpha_M$ as a function of $M$ is given by its $\beta$ function which is written as
\begin{equation}\label{betaM}
\beta_M\equiv M \frac{\partial \alpha_M(M)}{\partial M} \equiv \alpha_M^2 b_M(M).
\end{equation}
It involves three different contributions easily calculated from Eqs.~(\ref{piphi}-\ref{decou}).
The first one is associated to $\overline \Pi(M^2)$ and is equal, in the limit of large $M$, to $\beta_\eta$ as expected. The second one is associated to $\overline \Phi_1^{UV}(M^2)$ and has no equivalence in $\beta_\eta$. 
The third  one is associated to $\overline \Phi_1^{IR}(M^2)$, with also no equivalence in $\beta_\eta$. It is  $IR$ finite.
To get some insight into these contributions, let us  investigate $\beta_M$ in two different limits:
\begin{itemize}
\item In the limit of small energy scale, $M\ll m$, or equivalently in the limit of  heavy electron mass, we have 
\begin{equation}
\beta_M(M\ll m)=\frac{\alpha_M^2}{2\pi} \frac{19}{15}\frac{M^2}{m^2} +{\cal O}(\alpha_M^3).
\end{equation}
It  goes therefore to zero in the limit of infinitely large electron mass. This insures the decoupling of very heavy degrees-of-freedom ({\it d.o.f.}) from light ones, as expected from a mass-dependent $RG$ equation.
\item In the limit of large momentum scale, $M\gg m$, we get
\begin{equation} \label{bmm}
\beta_M(M\gg m)= -\frac{2 \alpha_M^2}{\pi} \mbox{Log}\frac{M^2}{m^2}+{\cal O}(\alpha_M^3).
\end{equation}
Remarkably enough, the $\beta_M$ function in this limit is negative and mass-dependent. It is discussed in more details below.
\end{itemize}

{\it iii)} The above decoupling equation between $\eta$- and $M$-dependences is also important in order to understand how the requirement for a perturbative calculation to remain valid should be understood. The only relevant (physical) coupling constant is $\alpha_M$, expressed in terms of the physical parameter $M$. This coupling constant  should be small compared to $1$ in order to be able to perform a meaningful perturbative calculation. This constraint, however, does not imply any  constraint on $\eta$ since the behavior of $\alpha_0$ as a function of $\eta$ is decoupled from the behavior of $\alpha_M$ as a function of $M$. In other words, $\eta$ can be chosen in principle to be very large, with  $\alpha_0(\eta)$ also very large, while maintaining a well defined perturbative calculation in terms of $\alpha_M$. From a practical point of view however, $\eta$ should  be chosen in order to avoid large numerical cancellations between $\alpha_0(\eta)$ and terms explicitly dependent on $\eta$, as shown in Eq.~(\ref{decou}), order by order in perturbation theory. This argument translates also to $DR$ with the identification $\mu=\eta m$ \cite{jfm}. We emphasize again that this particular choice of $\eta$, or equivalently of $\mu$, should not lead to any physical interpretation.

\subsection{Asymptotic behavior} \label{AF}
\begin{figure}
\centerline{\includegraphics[width=18pc]{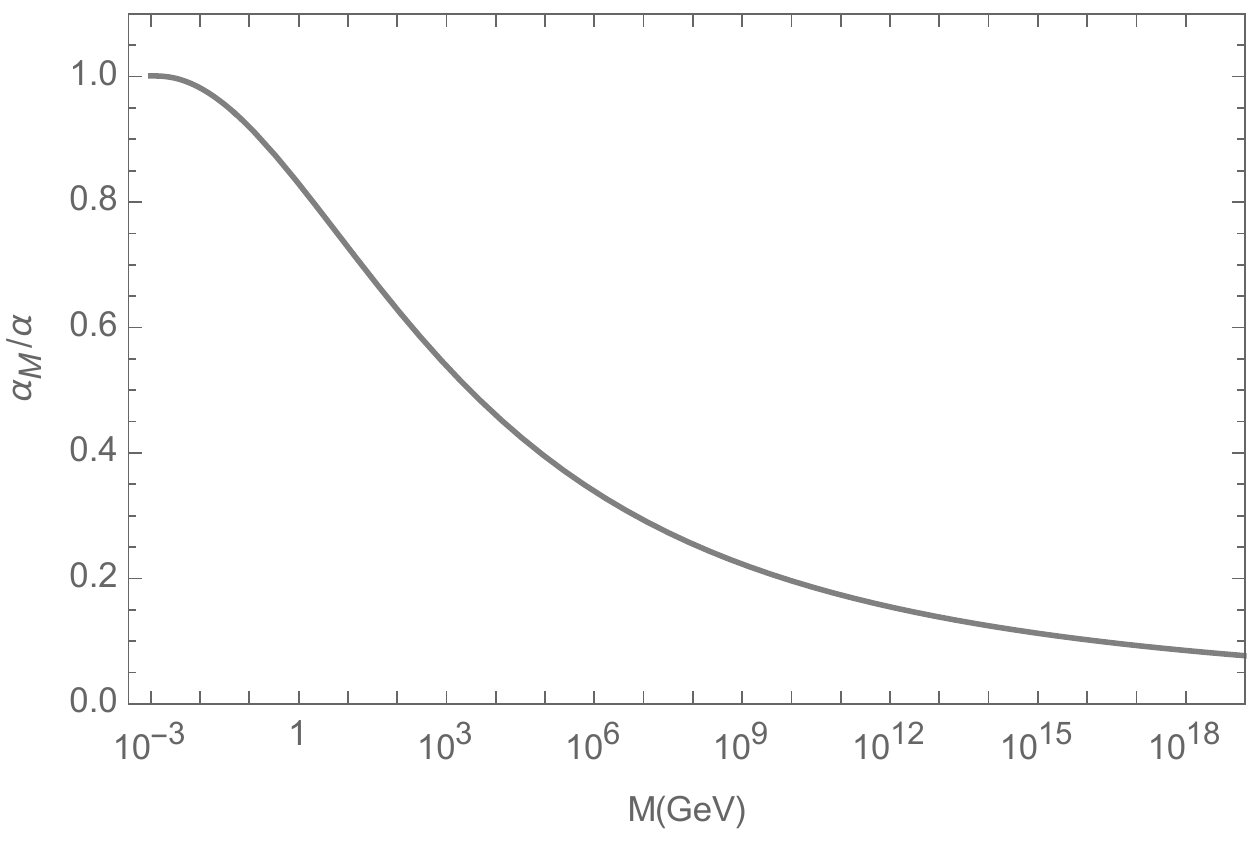}}
\begin{center}
\caption{The effective charge $\alpha_M$ divided by the fine structure constant - or equivalently the square of the $F_1$ form factor of the electron - as a function of $M$.}
\label{figalpha}
\end{center}
\end{figure}
\begin{figure}
\centerline{\includegraphics[width=16pc]{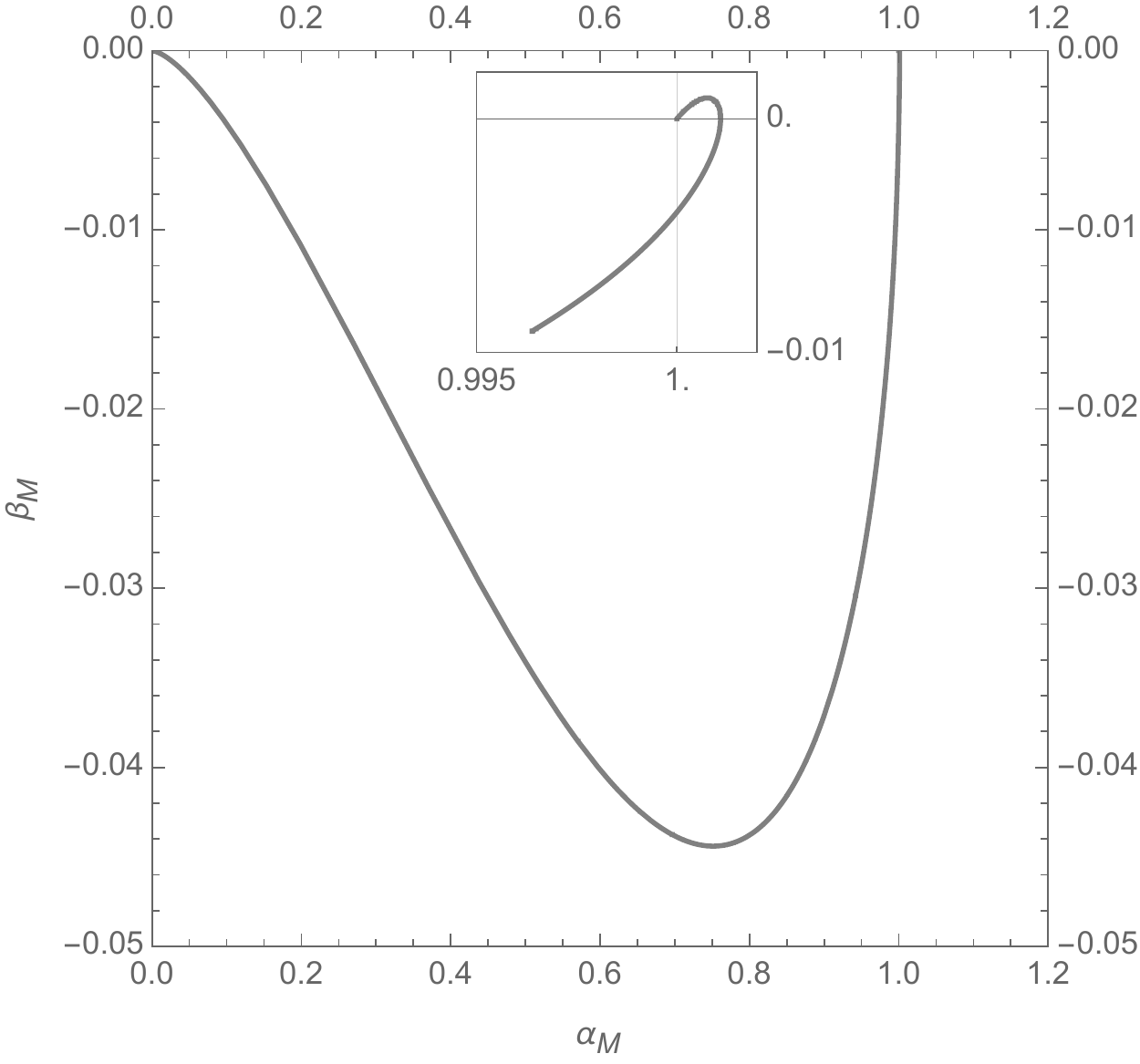}}
\begin{center}
\caption{The $\beta_M$ function as a function of $\alpha_M$, showing an ultra-violet stable fixed point at $\alpha^*=0$. The inserted figure shows a zoom on $\beta_M$ for $M\le 50$ GeV.}
\label{figbeta}
\end{center}
\end{figure}

From integration of the $\beta_M$ function of Eq.~(\ref{betaM}), we immediately get
\begin{equation}
\alpha_M(M)=\frac{\alpha}{1-\alpha \int_0^M d\nu \frac{b_M(\nu)}{\nu}},
\end{equation}
where $b_M$ is defined in Eq.~(\ref{betaM}). This effective charge is indicated on Fig.~\ref{figalpha}. It shows two immediate and far reaching consequences.
\begin{itemize}
\item The physical coupling constant does not show any Landau pole. This is at variance with the bare coupling constant which, as expected from Eq.~(\ref{betaeta}), exhibits a Landau pole at a critical value of the regularization scale $\eta$. As already emphasized in Ref.~\cite{jfm}, this Landau pole for $\alpha_0(\eta)$ should not have any physical interpretation.
\item The physical coupling constant shows an asymptotic freedom type behavior at very large energies. This is a direct consequence of the mass-dependent  contribution to $\beta_M$ at large $M^2$ originating from the taming of $IR$ divergences in $TLRS$ for the calculation of the electromagnetic vertex function.
\end{itemize}
The corresponding $\beta_M$ function is indicated on Fig.~\ref{figbeta}. As expected, it exhibits both an $IR$ stable fixed point at $\alpha_M(M=0)=\alpha$, and an $UV$ stable fixed point at $\alpha_M(M\to \infty)=0\equiv\alpha^*$. Note that for the calculation of the physical coupling constant, the limit of high energy scale is identical to the limit of small electron mass. This implies immediately that the physical coupling constant at finite $M$ should tend to zero when the electron mass tends to zero.
%

%%%%%%%%%%%%%%%%%%%%%%%%%%%%%%%%%%%%%%%%%%%%%%%%
\subsection{The case of two mass scales} \label{2mass}
To complete our discussion, let us consider  the case of two fermionic degrees of freedom, one with a low mass $m_<$, the other one with a high mass $m_>$, with the hierarchy $m_< \ll m_>$. We shall concentrate for this discussion on the bare and physical coupling constants. 

The running of the bare coupling constant is entirely given  by the vacuum polarization of the photon. It is mass independent, as recalled above. The only change when considering these two {\it d.o.f.} is thus just to multiply the $\beta_\eta$ function by two, without any consideration of threshold effect. The running of the bare coupling constant just counts the number of (charged)  {\it d.o.f.} present in the Lagrangian we start from, independently of their mass. Since any physical observable is independent of the regularization scale $\eta$, this change of the $\eta$ dependence of the bare coupling constant has absolutely no influence on the calculation of physical observables.

We should thus concentrate only on the behavior of the physical coupling constants as a function of the physical energy scale $M$, or in other words on the behavior of the electromagnetic form factors for the light or heavy {\it d.o.f.}. We denote by $\alpha_M^<$ and $\alpha_M^>$ the physical coupling constants of the light and heavy {\it d.o.f.} respectively.
The only new contribution to consider as compared to the calculation of the electromagnetic form factor in Sec.~\ref{phys}, with $m=m_<$ or $m=m_>$, corresponds to the contribution of the vacuum polarization of the photon.  We can identify three characteristic  kinematical conditions:

\begin{itemize}
\item $M \ll m_< \ll m_>$ \\ Since the vacuum polarization of the photon behaves in this limit like $M^2/m^2$, both contributions from light or heavy {\it d.o.f.} are negligible and the corresponding form factors are  close to 1. This insures that both coupling constants $\alpha_M^<$ and $\alpha_M^>$ are  equal to $\alpha$ for $M=0$, as they should.
\item $m_< \ll m_> \ll M$ \\ In this case, the contribution from the vacuum polarization of the photon is given by
\begin{equation}
\frac{\alpha_M^<}{6 \pi} \mbox{Log }\frac{M^2}{m_<^2} + \frac{\alpha_M^>}{6 \pi} \mbox{Log} \frac{M^2}{m_>^2}.
\end{equation}
This contribution is however subdominant as compared to the $IR$ contributions in the large $M$ kinematical region, as detailed in the preceeding subsection. 
\item $m_< \ll M  \ll m_>$ \\ This case is very similar to the above one, since the contribution from the heavy {\it d.o.f.} behaves in this condition like $M^2/m_>^2$ which is subdominant.
\end{itemize}
According to the above discussion, the physical coupling constant $\alpha_M^<$  is thus almost identical to the physical coupling constant with  one mass scale only, as discussed in the preceding sections.  This is in complete agreement with the decoupling theorem \cite{appel}. On the contrary, the physical coupling constant $\alpha_M^>$ is almost equal to $\alpha$ except in the very far $UV$ domain $M\gg m_>$ where it tends to zero like $\left[\mbox{Log}\frac{M^2}{m_>^2}\right]^{-1}$.

%%%%%%%%%%%%%%%%%%%%%%%%%%%%%%%%%%%%%%%%%%%%%%%%
\subsection{The limit of massless electrons} \label{zero}
The discussion of the limit of massless electrons, and the appearance of associated $IR$ divergences,  is usually done in terms of exceptional or non-exceptional momenta \cite{weinb}. Momenta are said exceptional if any partial sum of external momenta is zero. This classification, however, does not make any reference to whether the amplitude under consideration is a physical one or not. It should thus be clarified. 

Let us first recall the three different objects we have to manipulate in any calculation of a cross-section. An {\it elementary amplitude} is a single diagram which contributes to this cross-section, as investigated for instance in Sec.~\ref{elm}. Its calculation does not make any a-priori  assumption about the external legs. When these legs correspond to physical, real, particles, they are on their mass-shell: the corresponding amplitude is thus called a {\it physical amplitude}. Finally, a {\it physical observable} corresponds to the sum of all physical amplitudes contributing to a physical process at a given order of perturbation theory.

From the above classification, it is clear that the relevant amplitudes to worry about when calculating a physical process are therefore physical amplitudes and not elementary ones. This will be our guiding line for discussing the  limit of massless electrons. 
What happens, however, for elementary amplitudes with off-shell external legs? This case corresponds to the calculation of diagrams beyond $NLO$ in perturbation theory. The off-shell self-energy $\Sigma(p)$ for instance will  be attached in this case to an internal line of a more complex physical amplitude. For this more complex amplitude, the external legs of $\Sigma(p)$ contribute to internal propagators with appropriate test functions, according to the use of $TLRS$.  These additional test functions will prevent any new $UV$ as well as $IR$ singularities in such a way that only the final, on-shell, physical observable is independent of the regularization scale $\eta$ without any  new $IR$ singularities. The (apparent) singularities appearing for exceptional momenta will thus be taken care of in $TLRS$ thanks to the presence of the test functions in all internal propagators.\\

According to our discussion in Sec~\ref{RGE}, the calculation of  the massless limit of the electron for the different $\beta$ functions is immediate. On the one hand, the case of $\beta_\eta$ is trivial but also particularly instructive. Since it is mass-independent, its value for a massless electron is given by Eq.~(\ref{betaeta}). It is therefore non-zero, with the bare coupling constant given by (\ref{decou}). This would not be the case, however, if the regularization scale $\eta$ would have been dimensionful, as it is the case  in the standard formulation of $DR$ for instance. In this scheme indeed, the regularization scale is identified with the unit of mass \cite{thoo} $\mu$ of $DR$. In  absence of any other mass scale at the level of the $QED$ Lagrangian, the dimensionless renormalized coupling constant in the $MS$ scheme cannot depend on a single dimensionful variable only. It should therefore be a constant independent of $\mu$. This would imply a zero $\beta$ function for $QED$ at $NLO$, in contradiction with the mass-independence of the (non-zero) $\beta$ function in $DR+MS$.

On the other hand, the case of $\beta_M$ is  not trivial. Since $\alpha_M$ depends on the dimensionful variable $M$ through the ratio $M^2/m^2$, it should be independent of $M$ in the limit of massless electrons for obvious dimensional arguments. Since this limit is also equivalent to the large $M$ limit,  this implies that, by construction, one should have an $UV$ stable fixed point at a given value $\alpha^*$ of the physical coupling constant. This is precisely what we get from the analysis of the $\beta_M$ function in Sec.~\ref{AF}, with $\alpha^*=0$. In  the limit of massless electrons, the physical coupling constant at finite $M$ should therefore also tend to zero.  Note that this is only true in the absence of any other mass scale in the physical world, {\it i.e.} in  absence of spontaneous symmetry breaking. This is not true for instance for quantum chromodynamics ($QCD$).

%%%%%%%%%%%%%%%%%%%%%%%%%%%%%%%%%%%%%%%%%%%%%%%%
\section{Conclusions} \label{conc}
We have reanalyzed in this study first order radiative corrections to $QED$ in the light of the recently proposed regularization procedure called $TLRS$. While these corrections are by now standard textbook exercises, the use of $TLRS$ enables a completely new insight into our understanding of quantum corrections: it enables a direct calculation of physical observables in physical conditions starting from the bare Lagrangian itself, without encountering any $UV$  nor $IR$ divergences. The only renormalization we should worry about is the finite, physical, renormalization needed to calculate physical parameters in terms of bare ones.
These  features refer to the unique properties of $TLRS$: as an {\it a-priori} regularization procedure it enables us to give a mathematically well defined meaning to the local Lagrangian we start from, and as an {\it intrinsic} regularization procedure, all intermediate calculations are done in physical conditions, {\it i.e.} in four space-time  dimensions with massless photons, as required by gauge invariance.

The analysis of any physical observable should  be done in terms of two different sets of parameters: the bare coupling constants and bare masses defined at the level of the bare Lagrangian, and the physical ones - physical coupling constants and physical masses - which are measured experimentally and which are used to fix the value of the bare parameters. These two sets of parameters do depend on two different running variables. 
On the one hand, the bare coupling constant depends on the regularization scale $\eta$. This scale is inherent to the local character of the Lagrangian we start from, which is constructed from the product of fields or derivative of fields at the same space-time point. It is therefore associated to the scaling invariance of the $UV$ limit since for any internal momentum $k\to \infty $, we also have $\eta~ k\to \infty$, for any  finite scaling variable $\eta$. This regularization scale is therefore dimensionless. This is at variance with the usual dimensionful unit of mass of $DR$ for instance. Note that using $DR$ we can also identify a corresponding dimensionless variable, as explained in Refs~\cite{jfm,GW_gauge}. On the other hand, the physical coupling constant depends on the kinematical conditions chosen to measure it. This is the so-called renormalization point $M$. By construction, $M$ is dimensionful. 

The relationship between these two coupling constants, $\alpha_0$ and $\alpha_M$, is governed by a decoupling equation, as given by Eq.~(\ref{decou}). This equation can also be used to understand how the decoupling between heavy and light {\it d.o.f.} is at work. From its mass-dependence, the physical coupling constant  exhibits explicitly the decoupling between these {\it d.o.f.}, as well known. This behavior however is in complete agreement  with the mass-independence of the $\eta$-dependence of $\alpha_0$, as shown by this decoupling equation. This is indeed expected from general physical arguments since $\eta$ is associated to the scaling properties of elementary amplitudes in the  $UV$ regime. These ones are therefore  the same for any {\it d.o.f.} of finite mass, light or heavy. It is just associated to the local character of the Lagrangian we start from. As a direct consequence, the running of the bare coupling constant as a function of $\eta$ should thus include {\it all} (charged) physical {\it d.o.f.} present in the Lagrangian we start from. This should also be the case for the renormalized coupling constant in $DR$: the running of the renormalized coupling constant in $DR+MS$ should  include {\it all} (charged) physical {\it d.o.f.} present in the Lagrangian of the Standard Model, with no threshold effects depending on the energy scale under consideration.

Physical observables, when calculated in $TLRS$, are   free from any $IR$ divergences. This is the case for the  electromagnetic form factors of the electron which can be calculated directly with massless photons. The $F_1$ form factor is then independent of the characteristics of the detector and of its ability to discriminate  a single electron from an electron-photon state. This unique property of $TLRS$ opens the way for a direct unambiguous measurement of this form factor and at the same time could provide a direct non-trivial test of the validity of $TLRS$. The calculation of this form factor may also have important consequences for precision experiments involving $QED$, like for instance the calculation of the charge radius of the proton from electron scattering experiments \cite{gao}. It also immediately implies the presence of an $UV$ stable fixed point for the physical coupling constant at $\alpha^*=0$. Note that the $UV$ behavior of the physical coupling constant is entirely dominated by the behavior of the vertex function in the $IR$ domain, and is thus $UV$ complete. 

This finding has remarkable consequences. It implies both the absence of a Landau pole for the physical coupling constant, as well as an asymptotic freedom type behavior. This $UV$ fixed point at  $\alpha^*=0$ is also required by dimensional analysis since the limit of very high energy scale is identical to the limit of  zero electron mass. In this limit, the physical coupling constant  - which is dimensionless - should be independent of any energy scale, hence its $\beta$ function should be zero. This implies therefore  from Eq.~(\ref{bmm}) that $\alpha_M$ should be $0$ in this limit, in first order perturbation theory.  This may explain why the fine structure constant is small, since the electron mass is also small. The physical coupling constant  $\alpha_M$ at very large energy scale $M$ thus shows  an asymptotic freedom type behavior, similarly to the effective coupling constant of $QCD$ extracted for instance from the Bjorken sum rule \cite{deur}. This behavior should completely change our understanding of a possible unification of the physical coupling constants at very high energy \cite{amald}.

%%%%%%%%%%%%%%%%%%%%%%%%%%%%%%%%%%%%%%%%%%%%%%%%
\section* {Acknowledgement}
Preprint of an article published in Int. J. Mod. Phys. A,   2250204 (2022), doi:10.1142/S0217751X22502049 © World Scientific Publishing Company.

%%%%%%%%%%%%%%%%%%%%%%%%%%%%%%%%%%%%%%%%%%%%%%%%%%%%
\appendix
\section{Main properties of $TLRS$} \label{TLRS}
The general mathematical properties of $TLRS$ have been detailed in Refs.~\cite{GW} and \cite{GW_gauge}. Several applications have already been considered: application to light-front dynamics \cite{LFD}, interpretation of the fine-tuning of the Higgs mass in the Standard Model \cite{Higgs}, the recovery of the axial anomaly \cite{axial}, the fate of the trace anomaly \cite{trace} or the study of conformal field theories in two-dimensions \cite{GWm}. For completness, we shall briefly recall in this Appendix the main properties of $TLRS$. 

As already mentioned in the Introduction, this procedure originates from the well known observation that the divergencies of bare amplitudes can be traced back to the violation of causality due to ill-defined products of distributions at the same point \cite{bogol,bosh}. It requires therefore the whole apparatus of the theory of distribution \cite{schwa} to correctly define any local Lagrangian. We consider here for simplicity the case of a scalar field.

As detailed in Ref.~\cite{GW}, the physical field $\varphi$ is constructed in $TLRS$ as a functional of the original quantum field $\phi(x)$, considered as a distribution, according to
\begin{equation} \label{distri}
\varphi[\rho](x)\equiv \int d^4 y \phi(y) \rho(x-y),
\end{equation}
where the reflection symmetric test function $\rho$ belongs to the Schwartz space  $\mathscr{S}$ of rapidly decreasing functions \cite{schwa}. 
The physical interest to use the test function $\rho$ is to smear out the original distribution in a  space-time domain of typical extension $a$. The test function can thus be characterized by $\rho_a(x)$ and the physical field by $\varphi_a(x)\equiv \varphi[\rho_a](x)$. 

For practical calculations, it is convenient to construct the physical fields in momentum space. If we denote by $f_\sigma$ the Fourier transform of the test function $\rho_a$, we can write ${\varphi_a}$ in terms of creation and annihilation operators, leading to \cite{GW}
\begin{equation} \label{fk}
\varphi_a(x)\!=\!\!\int\!\frac{d^3{\bf p}}{(2\pi)^3}\frac{f_\sigma(\varepsilon_p^2,{\bf p}^2)}{2\varepsilon_p}
\left[a^\dagger_{\bf p} e^{i{p.x}}+a_{\bf p}e^{-i{p.x}}\right],
\end{equation}
with  $\varepsilon^2_p = {\bf p}^2+m^2$.  
Each propagator being the contraction of two fields is proportional to $f_\sigma^2$. 
This test function in momentum space is a dimensionless quantity. It should therefore be expressed in terms of dimensionless arguments. To do that, one shall introduce an arbitrary scale $M_0$ to "measure" all momenta. In practical calculations, $M_0$ can be any of the non-zero physical mass of the theory under consideration \footnote{For purely massless theories, $M_0$  corresponds for instance to the scale fixed to measure any non-zero momentum.}. It is taken equal to the electron mass $m$ in this study. Note that a change in the value of $M_0$ is just equivalent to a redefinition of $\eta$, without any consequences on physical observables which should anyhow be independent of $\eta$ and $M_0$. 

As shown in Ref.~\cite{GW}, it is appropriate to choose $f_\sigma$  as a partition of unity.  A simple example of such function, with $f_\sigma$ constructed from the sum of two elementary functions only, is discussed in Ref.~\cite{LFD}. It  is equal to $1$ almost everywhere and is $0$ outside a finite domain of $\reels^{+ 4}$, along with all its derivatives (super-regular function).
The parameter $\sigma$, chosen for convenience between 0 and 1, controls the lower and upper limits of the support of $f_\sigma$.  Note that for any partition of unity, the product of two partitions of unity is also a partition of unity. We shall therefore identify $f_\sigma^2$ by $f_\sigma$ when needed.  As we shall see in \ref{integral}, we do not need to know the precise form of the test function as a partition of unity, we just rely on its asymptotic properties. Note that the construction of the test function as a partition of unity is essential in order to relate its $IR$ and $UV$ properties.

Requiring locality for the bare Lagrangian we start from  implies considering the subsequent  limit $a \to 0$, dubbed {\it the continuum limit.} In this process, it is essential to preserve the scaling properties
\begin{equation} \label{scale}
\rho_a(x) \to^{^{\!\!\!\!\!\!\!\!\!a\to 0}} \rho_\eta(x) \ ; \  \varphi_a(x) \to^{^{\!\!\!\!\!\!\!\!\!a\to 0}} \varphi_\eta(x),
\end{equation}
where $\eta$ is an arbitrary, dimensionless, scaling variable since in the limit $a \to 0$, we also have $a/\eta \to 0$, for any finite $\eta$. This arbitrary scale just governs the  ``spead'' at which the continuum limit is reached. Any physical observable should of course be independent of this dimensionless scaling variable, also called regularization scale in order to stick to the common denomination, although this denomination may be misleading when using $TLRS$ since this regularization scale is dimensionless in this case. From the choice of parametrization of $f_\sigma$, the continuum limit corresponds to $\sigma \to 1^-$.

As we shall see in the next Appendix, any amplitude associated to a singular operator $T(X)$ is written schematically as
\begin{equation} \label{cala}
{\cal A}_\sigma= \int_0^\infty dX \ T(X) \ f_{\sigma}(X).
\end{equation}
We consider here for simplicity a one-loop amplitude, with a single dimensionless variable $X$.
It is easy to check that a na\" ive implementation of the continuum limit for the test function, with a constant boundary condition like for instance $X \le H_\sigma=1/(1-\sigma)$, will result in a divergent amplitude $\lim_{\sigma \to1^-}A_\sigma$, as expected from the calculation of the amplitude in terms of a cut-off in momentum space. However, from the scaling properties (\ref{scale}), we should get
\begin{equation} \label{Aeta}
A_\eta \equiv \lim_{\sigma \to1^-}A_\sigma.
\end{equation}
To achieve this, we should rather consider a ``running'' boundary condition on $f_\sigma$ defined by
\begin{equation} \label{running}
f_\sigma(X \ge H_\sigma(X)) = 0,
\end{equation}
with
\begin{equation} \label{scaling}
H_\sigma(X)\equiv \eta^2 X g_\sigma(X) + (\sigma - 1),
\end{equation}
where $\eta$ is the dimensionless regularization scale in $TLRS$ \footnote{The square of $\eta$ in (\ref{scaling}) is just for convenience since $X$ is usually identified with the square of a momentum, as shown in \ref{integral}.} with $\eta^2>1$.  The function $g_\sigma(X)$ is constructed in such a way that the boundary on $X$, as defined by Eq.~(\ref{running}), is finite and tends to $1^-$ when $\sigma \to 1^-$.  A typical (but not unique) simple form for $g_\sigma(X)$ is given by
\begin{equation}
g_\sigma(X)=X^{\sigma-1}.
\end{equation}
The conditions (\ref{running}) and (\ref{scaling})  amount to an infinitesimal drop-off of the test function in the $UV$ region, with the drop-off rate governed by the regularization scale $\eta$. 
It also preserves  the super-regular properties of the test function in the continuum limit, with the test function and all its derivatives being zero at infinity.

Remarkably enough, this boundary condition defines at the same time the $UV$ and $IR$ boundaries once $f$ is constructed from a partition of unity \cite{LFD}. The explicit calculation of standard one-loop integrals using $TLRS$ is thus straightforward, as recalled in \ref{integral}. It relies on the identification of the test function $f_\sigma$ with its Taylor remainder in the $UV$ as well as $IR$ domains - thanks to its asymptotic properties -  and the subsequent use of the Lagrange formula, hence the name Taylor-Lagrange regularization scheme \cite{GW}.  The calculation of elementary amplitudes in the $UV$ and $IR$ domains is thus immediate. In the $UV$ domain the continuum limit (\ref{Aeta}) is  taken after  integration by part. The extension of $IR$ singular operators \cite{GW,GW_gauge} involves the Pseudo-function \cite{schwa,GWm}, denoted by $Pf$, of $1/X^n$, with $n\ge 1$. This gives, for $n>1$,
\begin{equation}
\int_0^{X_0} dX \mbox{Pf}\left[\frac{1}{X^n}\right]=\lim_{\epsilon\to 0}\left[\int_\epsilon^{X_0} \frac{dX}{X^n}-\frac{1}{1-n}\frac{1}{\epsilon^{n-1}}\right],
\end{equation}
and for $n=1$
\begin{equation}
\int_0^{X_0} dX \mbox{Pf}\left[\frac{1}{X}\right]=\lim_{\epsilon\to 0}\left[\int_{\lambda \epsilon}^{X_0} \frac{dX}{X}+\mbox{Log}( \epsilon)\right],
\end{equation}
where $\lambda$ is an arbitrary scale variable \cite{schwa}. The value of $\lambda$ is fixed from the choice of gauge \cite{GW_gauge}. In the Feynman gauge we have $\lambda =1$.

\section{Relevant integrals} \label{integral}
For completness, and as an illustration of the use of $TLRS$ in practical calculations, we detail in this Appendix all the relevant integrals needed in our study.

\subsection{Self-energy of the electron}
\subsubsection{Calculation of $\overline I_0$}
We recall here the various steps of the calculation of this simple integral. More details can be found in the Appendix of Ref.~\cite{axial}. We calculate the relevant integrals in Euclidian space, using Feynman representation. The integral $\overline I_0$  is written
\begin{equation}
\overline I_0(p)=\frac{i}{(2\pi)^4}\lim_{\sigma \to1^-}\int_0^1dx\int d^4{\bf K} \frac{1}{[{\bf K}^2+m^2 x \Delta_p]^2}F_\sigma,\nonumber
\end{equation}
where $F_\sigma$ is a simplified notation for the product of the two test functions, with
\begin{equation}
F_\sigma=f_\sigma\left[\frac{({\bf K}+x p)^2}{M_0^2}\right] f_\sigma\left[\frac{({\bf K}-(1-x)p)^2}{M_0^2}\right].\nonumber
\end{equation}
For a non zero electron mass, it is convenient to choose $M_0\equiv m$. In the absence of test functions, this integral is divergent in the $UV$ regime only. 
We can thus safely concentrate on the behavior of the test functions for large ${\bf K ^2}$. The use of test functions when $IR$ singular operators are involved is detailed in \ref{IRdiv}.
In the $UV$ domain, the arguments of the two test functions are both equivalent to ${\bf K}^2/m^2 \equiv \Delta_p X$ with $\Delta_p\neq 0$. We extract here from the running variable $X$ the scale $\Delta_p$ which depends on the kinematical conditions. This insures that the integrand $X/(X+x)^2$ is independent of any momentum-dependent scale so that the scaling variable $\eta$ is also (implicitely) independent of any momentum-dependent scale. We thus have, 
with the identification $f_\sigma^2 \sim f_\sigma$ valid for a partition of unity,
\begin{equation}
\overline I_0(p)=\frac{i}{(4\pi)^2}\lim_{\sigma \to1^-}\int_0^1dx\int_0^\infty dX \frac{X}{(X+x)^2}f_\sigma\left[\Delta_p X\right].\nonumber
\end{equation}
From the properties of the test function \cite{GW}, we can write a Lagrange formula for $f_\sigma$, at fixed support \cite{axial}, with
\begin{equation}
f_\sigma\left[\Delta_p X\right]=-X\int_{\Delta_p}^\infty \frac{dt}{t} \frac{\partial}{\partial X} f_\sigma\left[X t\right].\nonumber
\end{equation}
We can then write $\overline I_0$ in the following form, after  integration by part, \begin{equation}
\overline I_0(p)=\frac{i}{(4\pi)^2}\lim_{\sigma \to1^-}\int_0^1dx\int_0^\infty dX \frac{\partial}{\partial X}\left[ \frac{X^2}{(X+x)^2}\right]\int_{\Delta_p}^\infty \frac{dt}{t} f_\sigma[Xt] .\nonumber
\end{equation}
From the boundary condition (\ref{running}), the argument of $f_\sigma$ under the integral is bounded from above by the support of the test function given by $H_\sigma(X)$, so that
\begin{equation}
Xt\le \eta^2 X g_\sigma(X)\ \ \mbox{and}\ \ \Delta_p \le t\leq \eta^2g_\sigma(X).\nonumber
\end{equation}
Since the integral over $X$ is now finite thanks to the derivative, we can safely take the continuum limit $\sigma \to1^-$ which gives $g_\sigma\to 1$ and $f_\sigma \to 1$. 
We finally get
\begin{equation}
\overline I_0(p)=\frac{i}{(4\pi)^2}\int_0^1dx\ \mbox{Log}\frac{\eta^2}{\Delta_p}.\nonumber
\end{equation}
A different choice for $M_0$ will just induce a multiplicative factor at $\eta^2$. This shall induce a finite additive constant on top of any contribution in $\mbox{Log}\ \eta^2$, as indicated in the final results for the elementary amplitudes calculated in Sec.~(\ref{elm}), with no consequences for any physical observables. This is reminiscent of the flexibility in choosing the unit of mass $\mu$ of $DR$ \cite{thoo}, like for instance using either the $MS$ or the $\overline{MS}$ schemes.

\subsubsection{Calculation of $\overline I_1^\mu$}
The integral  $\overline I_1^\mu$ is written as
\begin{equation}
\overline I_1^\mu(p)=I_1^\mu(p)+{\bf p}^\mu \frac{i}{(2\pi)^4}\lim_{\sigma \to1^-}\int_0^1dx\int d^4{\bf K} \frac{x}{[{\bf K}^2+m^2 x \Delta_p]^2}F_\sigma,\nonumber
\end{equation}
with
\begin{equation}
I_1^\mu(p)=\frac{i}{(2\pi)^4}\lim_{\sigma \to1^-}\int_0^1dx\int d^4{\bf K} \frac{{\bf K}^\mu}{[{\bf K}^2+m^2 x \Delta_p]^2}F_\sigma,\nonumber
\end{equation}
The term in ${\bf p}^\mu$ is calculated similarly to $\overline I_0$. The calculation of $I_1^\mu$ should be done with care since the test functions do depend on all the relevant momenta of the system. Following the calculations of Ref.~\cite{axial}, we start from the identity
\begin{equation}
\frac{\partial}{\partial {\bf K}_\mu} \frac{1}{{\bf K}^2+m^2 x \Delta_p}=-2 \frac{{\bf K}^\mu}{({\bf K}^2+m^2 x \Delta_p)^2}.\nonumber
\end{equation}
We can thus write immediately
\begin{equation}
I_1^\mu=-\frac{i}{2(2\pi)^4}\lim_{\sigma \to1^-}\int_0^1dx\int \frac{d^4 {\bf K}}{(2\pi)^4}\frac{\partial}{\partial {\bf K}_\mu} \left[\frac{1}{{\bf K}^2+m^2 x \Delta_p} \right]F_\sigma.\nonumber
\end{equation}
By integration by part, the surface term is a 3-dimensional integral orthogonal to the $\mu$-direction. It should be taken at ${\bf K}_\mu \to \pm \infty$. Thanks to the presence of the test functions, this term is identically zero. The remaining integral involves the derivative of $F_\sigma$, with 
\begin{equation}
F_\sigma=f_\sigma\left[\frac{({\bf K}+x{\bf p})^2}{m^2}\right]f_\sigma\left[\frac{({\bf K}-(1-x){\bf p})^2}{m^2}\right].\nonumber
\end{equation}
One thus gets
\begin{eqnarray}
I_1^\mu=&&\frac{i}{m^2(2\pi)^4}\lim_{\sigma \to1^-}\int_0^1dx\int \frac{d^4 {\bf K}}{(2\pi)^4}\frac{1}{{\bf K}^2+m^2 x \Delta_p}\nonumber\\
\times&& \left[ ({\bf K}^\mu+x{\bf p}^\mu) f_\sigma^\prime \left[\frac{({\bf K}+x{\bf p})^2}{m^2}\right]f_\sigma\left[\frac{({\bf K}-(1-x){\bf p})^2}{m^2}\right]\right. \nonumber \\
+&&\left.({\bf K}^\mu-(1-x){\bf p}^\mu) f_\sigma \left[\frac{({\bf K}+x{\bf p})^2}{m^2}\right]f_\sigma^\prime\left[\frac{({\bf K}-(1-x){\bf p})^2}{m^2}\right]\right].\nonumber
\end{eqnarray}
In this equation $f^\prime_\alpha$ denotes $\frac{d}{dX} f_\sigma(X)$. The integral $I_1^\mu$ is a-priori non-zero only in the $UV$ region where $f^\prime\neq 0$. In this region, all test functions are equivalent to $f_\sigma\left[\frac{{\bf K}^2}{m^2}\right]$. By symmetry arguments, the integral over ${ \bf K}_\mu$ is  strictly zero and it remains to calculate
\begin{equation}
I_1^\mu=\frac{i}{(2\pi)^4}\frac{{\bf p}^\mu}{m^2} \lim_{\sigma \to1^-}\int_0^1dx\int \frac{d^4 {\bf K}}{(2\pi)^4}\\
\frac{2x-1}{{\bf K}^2+m^2 x \Delta_p}f_\sigma\left[\frac{{\bf K}^2}{m^2}\right]f_\sigma^\prime\left[\frac{{\bf K}^2}{m^2}\right].\nonumber
\end{equation}
With $\Delta_p X={\bf K}^2/m^2$ we have
\begin{equation}
I_1^\mu=\frac{i}{2(4\pi)^2}{\bf p}^\mu \lim_{\sigma \to1^-}\int_0^1dx(2x-1)\Delta_p \int_0^{X_{max}} dX \frac{X}{X+x}\left[f_\sigma^2(\Delta_p X)\right]^\prime.\nonumber
\end{equation}
By integration by part, we have
\begin{multline}
I_1^\mu=\frac{i}{2(4\pi)^2}{\bf p}^\mu\lim_{\sigma \to1^-}\int_0^1dx(2x-1) \Delta_p \\
\left[\left. \frac{X}{X+x } f_\sigma^2(X)\right \vert_0^\infty-\int_0^\infty dX \left[\frac{X}{X+x }\right]^\prime f_\sigma^2(\Delta_p X)\right].\nonumber
\end{multline}
Both contributions are finite in the absence of the test functions, so that we can  safely take the continuum limit $f_\sigma \to 1^-$ and we finally get
\begin{equation}
I_1^\mu=0. \nonumber
\end{equation}
We recover here rotational invariance. Note that this property is only true in the continuum limit. We thus have finally for $\overline I_1^\mu$
\begin{equation}
\overline I_1^\mu(p)=\frac{i}{(4\pi)^2}\ p^\mu\int_0^1dx\ x\ \mbox{Log}\frac{\eta^2}{\Delta_p}.\nonumber
\end{equation}

\subsection{Vacuum polarization of the photon}
The calculation of $\overline J_0$ and $\overline J_1^\mu$ is very similar to the calculation of $\overline I_0$ and $\overline I_1^\mu$ detailed above, with $\Delta_p$ replaced by $\Delta_q$. We thus get immediately
\begin{eqnarray}
\overline J_0(p)&=&\frac{i}{(4\pi)^2}\int_0^1dx\ \mbox{Log}\frac{\eta^2}{\Delta_q},\nonumber\\
\overline J_1^\mu(p)&=&\frac{i}{(4\pi)^2}\ p^\mu\int_0^1dx\ x\ \mbox{Log}\frac{\eta^2}{\Delta_q}.\nonumber
\end{eqnarray}
\subsubsection{Calculation of $\overline J_2$}
Following the calculation of $\overline I_0$, the integral $\overline J_2$ is written as
\begin{equation}
\overline J_2(p)=J_2(p) -p^2 \frac{i}{(2\pi)^4}\lim_{\sigma \to1^-}\int_0^1dx\int d^4{\bf K} \frac{x^2}{[{\bf K}^2+m^2\Delta_p]^2}F_\sigma,\nonumber
\end{equation}
with
\begin{equation}
J_2(p)=-\frac{i}{(2\pi)^4}\lim_{\sigma \to1^-}\int_0^1dx\int d^4{\bf K} \frac{{\bf K}^2}{[{\bf K}^2+m^2\Delta_p]^2}F_\sigma.\nonumber
\end{equation}
The term in $p^2$ is calculated similarly to $\overline I_0$, and we have for $J_2$
\begin{equation}
J_2(p)=-\frac{i}{(4\pi)^2}m^2\Delta_q\lim_{\sigma \to1^-}\int_0^1dx\int_0^\infty dX \frac{X^2}{(X+1)^2}f_\sigma\left[\Delta_p X\right].\nonumber
\end{equation}
The integrand over $X$ can  be written as
\begin{equation}
\frac{X^2}{(X+1)^2}=1-\frac{1}{X+1}-\frac{X}{(X+1)^2}.\nonumber
\end{equation}
It is easy to check that the contribution of the first term to $J_2$ is strictly zero, with
\begin{equation}
\lim_{\sigma \to1^-}\int_0^\infty dX f_\sigma\left[\Delta_p X\right]=\lim_{\sigma \to1^-}\int_0^\infty \frac{dY}{Y^2} f_\sigma\left[\frac{\Delta_p}{Y}\right]=\int_0^\infty dY\mbox{Pf}\left[\frac{1}{Y^2}\right]\equiv 0.\nonumber
\end{equation}
It remains
\begin{equation}
J_2(p)=\frac{i}{(4\pi)^2}m^2\Delta_q\lim_{\sigma \to1^-}\int_0^1dx\int_0^\infty dX\left[ \frac{1}{X+1}+\frac{X}{(X+1)^2}\right]f_\sigma\left[\Delta_p X\right].\nonumber
\end{equation}
Using the Lagrange formula for $f_\sigma$, and after integration by part, we get, in the continuum limit
\begin{equation}
J_2(p)=\frac{i}{(4\pi)^2}m^2\Delta_q\int_0^1dx\int_0^\infty dX\frac{\partial}{\partial X}\left[ \frac{X}{X+1}+\frac{X^2}{(X+1)^2}\right]\int_{\Delta_q}^{\eta^2} \frac{dt}{t},\nonumber
\end{equation}
so that
\begin{equation}
J_2(p)=\frac{2i}{(4\pi)^2}\ m^2\Delta_q\int_0^1dx\ \mbox{Log}\frac{\eta^2}{\Delta_q}.\nonumber
\end{equation}
We thus get
\begin{equation}
\overline J_2(p)=\frac{i}{(4\pi)^2}\ m^2\Delta_q\int_0^1dx\left (2+\frac{x^2}{\Delta_q}\frac{p^2}{m^2}\right) \mbox{Log}\frac{\eta^2}{\Delta_q}.\nonumber \nonumber
\end{equation}

\subsubsection{Calculation of $\overline J_2^{\mu\nu}$}
Following the calculation of $\overline I_0$ and $\overline I_1^\mu$, the integral $\overline J_2^{\mu\nu}$ is written as
\begin{equation}
\overline J_2^{\mu\nu} = J_2^{\mu\nu}+p^\mu p^\nu \frac{i}{(2\pi)^4}\lim_{\sigma \to1^-}\int_0^1dx \int d^4{\bf K} \frac{x^2}{[{\bf K}^2+m^2\Delta_p]^2}F_\sigma, \nonumber
\end{equation}
with
\begin{equation}
J_2^{\mu\nu} = \frac{i}{(2\pi)^4}\lim_{\sigma \to1^-}\int_0^1dx\int d^4{\bf K} \frac{{\bf K}^\mu{\bf K}^\nu}{[{\bf K}^2+m^2\Delta_p]^2}F_\sigma, \nonumber
\end{equation}
The term in $p^\mu p^\nu$ is calculated similarly to $\overline I_0$. From symmetry arguments, and in the absence of any external momentum in the continuum limit, we can write
\begin{equation}
J_2^{\mu\nu}=A g^{\mu\nu}.\nonumber
\end{equation}
By contraction with $g_{\mu\nu}$, we get immediately
\begin{equation} \label{kmunu}
A=\frac{1}{4} g_{\mu\nu} I_2^{\mu\nu}.\nonumber
\end{equation}
Note that, due to the presence of the test functions,  the contraction $g_{\mu\nu} I_2^{\mu\nu}$  {\it is not} a-priori equal to $\tilde J_2$ written as
\begin{equation}
\tilde J_2(p)=-\frac{i}{(2\pi)^4}\lim_{\sigma \to1^-}g_{\mu\nu}\int_0^1dx \int d^4{\bf K} \frac{{\bf K}^\mu{\bf K}^\nu}{[{\bf K}^2+m^2\Delta_p]^2}F_\sigma.\nonumber
\end{equation}
The difference, if any, should come from the asymptotic behavior of the test functions in the continuum limit. This prevents to reverse the order of taking the continuum limit $\sigma \to 1^-$ with the contraction by $g_{\mu \nu}$. It cannot therefore depend on any mass scale. Since $J_2$ and $J_2^{\mu\nu}$ have a dimension 
of a mass squared, this difference is thus zero. This is however not the case for the integrals $K_2$ and $K_2^{\mu\nu}$, as we shall see below, since these integrals are dimensionless. We thus get 
\begin{equation}
J_2^{\mu\nu}=\frac{1}{4}g^{\mu\nu} \tilde J_2. \nonumber
\end{equation}
where $\tilde J_2$ can be deduced easily from $J_2$ with
\begin{equation}
\tilde J_2(p)=\frac{2i}{(4\pi)^2}\ m^2\Delta_q\int_0^1dx\ \mbox{Log}\frac{\eta^2}{\Delta_q}.\nonumber
\end{equation}

\subsection{Electromagnetic vertex}
The integrals involved in the calculation of the electromagnetic form factor, with three propagators, have already been detailed in Ref.~\cite{axial} for the calculation of the triangular diagrams leading to the axial anomaly. 
\subsubsection{Calculation of $\overline K_0$}
The integral $\overline K_0$ is written as
\begin{equation}
\overline K_0(p,q)= \frac{2i}{(2\pi)^4} \lim_{\sigma \to1^-} \int_0^1dx\int_0^{1-x}dy \int d^4{\bf K} \frac{1}{({\bf K}^2+m^2 \Delta)^3}F_\sigma,\nonumber
\end{equation}
with
\begin{equation}
F_\sigma=f_\sigma\left[\frac{({\bf K+P})^2}{m^2}\right] f_\sigma\left[\frac{({\bf K+P-p'})^2}{m^2}\right] f_\sigma\left[\frac{({\bf K+P-p})^2}{m^2}\right],\nonumber
\end{equation}
and $P=x p'+y p$. Since this integral is finite, we can safely take the continuum limit with $f_\sigma \to 1$ and get
\begin{equation}
\overline K_0(p,q)=-\frac{i}{(4\pi)^2}\int_0^1dx\int_0^{1-x}dy  \frac{1}{\Delta}.\nonumber
\end{equation}

\subsubsection{Calculation of $\overline K_1^\lambda$}
The integral $\overline K_1^\lambda$ is written as
\begin{equation}
\overline K_1^\lambda(p,q)= K_1^\lambda(p,q)+\frac{2i}{(2\pi)^4} \lim_{\sigma \to1^-} \int_0^1dx\int_0^{1-x}dy \int d^4{\bf K} \frac{{\bf P}^\lambda}{({\bf K}^2+m^2 \Delta)^3}F_\sigma,\nonumber
\end{equation}
with
\begin{equation}
 K_1^\lambda(p,q)= \frac{2i}{(2\pi)^4} \lim_{\sigma \to1^-} \int_0^1dx\int_0^{1-x}dy \int d^4{\bf K} \frac{{\bf K}^\lambda}{({\bf K}^2+m^2 \Delta)^3}F_\sigma,\nonumber
\end{equation}
In the continuum limit, ${K_1}^\lambda$ is strictly zero as shown in Ref.~\cite{axial} so that we get immediately, from the calculation of $\overline K_0$,
\begin{equation}
K_1^\lambda(p,q)=-\frac{i}{(4\pi)^2}\int_0^1dx\int_0^{1-x}dy  \frac{{\bf P}^\lambda}{m^2 \Delta}.\nonumber
\end{equation}

\subsubsection{Calculation of $\overline K_2$}
Similarly, the integral $\overline K_2$ is written as
\begin{equation}
\overline K_2(p,q)= K_2(p,q)+\frac{2i}{(2\pi)^4} \lim_{\sigma \to1^-} \int_0^1dx\int_0^{1-x}dy \int d^4{\bf K} \frac{{\bf P}^2}{({\bf K}^2+m^2 \Delta)^3}F_\sigma,\nonumber
\end{equation}
with
\begin{equation}
K_2(p,q)= \frac{2i}{(2\pi)^4} \lim_{\sigma \to1^-} \int_0^1dx\int_0^{1-x}dy \int d^4{\bf K} \frac{{\bf K}^2}{({\bf K}^2+m^2 \Delta)^3}F_\sigma,\nonumber
\end{equation}
From the results of Ref.~\cite{axial} we get immediately
\begin{equation}
\overline K_2(p,q)=\frac{i}{(4\pi)^2}\int_0^1dx\int_0^{1-x}dy \left[2\mbox{Log}\frac{\eta^2}{\Delta} -\frac{{\bf P}^2}{m^2 \Delta}\right].\nonumber
\end{equation}

\subsubsection{Calculation of $\overline K_2^{\mu\nu}$}
\begin{equation}
\overline K_2^{\mu\nu}(p,q)= K_2^{\mu\nu}(p,q)+
\frac{i}{(2\pi)^4} \lim_{\sigma \to1^-} \int_0^1dx\int_0^{1-x}dy \int d^4{\bf K} \frac{{\bf P}^\mu {\bf P}^\nu}{({\bf K}^2+m^2 \Delta)^3}F_\sigma,\nonumber
\end{equation}
with
\begin{equation}
K_2^{\mu\nu}(p,q)= 
\frac{i}{(2\pi)^4} \lim_{\sigma \to1^-} \int_0^1dx\int_0^{1-x}dy \int d^4{\bf K} \frac{{\bf K}^\mu {\bf K}^\nu}{({\bf K}^2+m^2 \Delta)^3}F_\sigma.\nonumber
\end{equation}
Following the above discussion for the calculation of $J_2^{\mu\nu}$, and according to the results of Ref.~\cite{axial}, we have
\begin{equation}
K_2^{\mu\nu}(p,q)=\frac{1}{4}g^{\mu\nu}\left[K_2(p,q)+m^2 \Delta \overline K_0(p,q)\right].\nonumber
\end{equation}

\subsection{Infra-red divergences} \label{IRdiv}
\subsubsection{Calculation of $A'$ and $B'$}
The calculation of $A'(m^2)$ and $B'(m^2)$ involves a singular integral in $1/x$. This singularity corresponds to a pole at ${\bf K}=0$. In this kinematical domain, the relevant test function is written as $f_\sigma\left(\frac{x^2 p^2}{m^2}\right)$. We thus have, for $p^2=m^2$,
\begin{eqnarray}
A'(m^2)&=&\frac{\alpha}{\pi m} \lim_{\sigma \to 1^-}\int_0^1 dx \frac{1-x}{x} f_\sigma(x^2)=\frac{\alpha}{\pi m}\left[\frac{1}{2}\int_0^1dX\mbox{Pf}\left(\frac{1}{X}\right)-1\right] \nonumber\\
&=&-\frac{\alpha}{\pi m},\nonumber
\end{eqnarray}
and
\begin{equation}
B'(m^2)=-\frac{\alpha}{2\pi m} \lim_{\sigma \to 1^-}\int_0^1 dx \frac{(1-x)^2}{x} f_\sigma(x^2)=\frac{3\alpha}{4\pi m}.\nonumber
\end{equation}

\subsubsection{Calculation of $\Phi_1^{IR}$}
The $IR$ singularities in the calculation of $\Phi_1^{IR}$ originates from the poles in $w=x+y=0$, {\it i.e.} for $x=y=0$. This pole occurs for ${\bf K}=0$ and is taken care of by the test function $f_\sigma\left[\frac{k^2}{m^2}\right]$. This test function is written as, in this limit,
\begin{equation}
f_\sigma\left[\frac{k^2}{m^2}\right]\to f_\sigma\left[\frac{{\bf P}^2}{m^2}\right]=f_\sigma\left[w^2\left(1+\frac{Q^2}{m^2}\xi(1-\xi)\right)\right], \nonumber
\end{equation}
with the variables $w$ and $\xi$ introduced in Eq.~(\ref{wxi}). The a-priori singular part of the relevant integrals is thus written as
\begin{eqnarray}
I&=&\lim_{\sigma \to1^-}\int_0^1\frac{dw}{w}f_\sigma\left(w^2\Delta_q\right)=\frac{1}{2}\lim_{\sigma \to1^-}\int_0^{\Delta_q}\frac{dX}{X}f_\sigma(X)=\frac{1}{2}\int_0^{\Delta_q}dX\mbox{Pf}\left(\frac{1}{X}\right)\nonumber \\
&=&\frac{1}{2}\mbox{Log}\Delta_q. \nonumber
\end{eqnarray}
Thanks to the presence of the test function, this contribution is finite, eventhough we have considered a massless photon. In a calculation using $DR$ with a finite mass $\delta$ of the photon, such integral will have an additional contribution in $\mbox{Log}\left(\frac{\delta^2}{m^2}\right)$.

%%%%%%%%%%%%%%%%%%%%%%%%%%%%%%%%%%%%%%%%%%%%%%%%

\end{document}